# Radiative Decay Width of Neutral non-Strange Baryons from PWA

Igor I. Strakovsky[1,a], William J. Briscoe[1], Alexander E. Kudryavtsev[1,2], Viacheslav V. Kulikov[2], Maxim A. Martemyanov[2], and Vladimir E. Tarasov[2]

[1] The George Washington University, Washington, DC 20052, USA
[2] Institute for Theoretical and Experimental Physics, Moscow, 117218, Russia

**Abstract.** An overview of the GW SAID and ITEP groups effort to analyze pion photoproduction on the neutron-target will be given. The disentanglement the isoscalar and isovector EM couplings of N* and Δ* resonances does require compatible data on both proton and neutron targets. The final-state interaction plays a critical role in the state-of-the-art analysis in extraction of the γn→πN data from the deuteron target experiments. It is important component of the current JLab, MAMI-C, SPring-8, ELSA, and ELPH programs.

## 1 Introduction

The N* family of nucleon resonances has many well established members [1], several of which exhibit overlapping resonances with very similar masses and widths but with different $J^P$ spin-parity values. Apart from the N(1535)1/2⁻ state, the known proton and neutron photo-decay amplitudes have been determined from analyses of single-pion photoproduction. The present work reviews the region from the threshold to the upper limit of the SAID analyses, which is CM energy W = 2.5 GeV. There are two closely spaced states above Δ(1232)3/2⁺: N(1520)3/2⁻ and N(1535)1/2⁻. Up to W ≈ 1800 MeV, this region also encompasses a sequence of six overlapping states: N(1650)1/2⁻, N(1675)5/2⁻, N(1680)5/2⁺, N(1700)3/2⁻, N(1710)1/2⁺, and N(1720)3/2⁺.

One critical issue in the study of meson photoproduction on the nucleon comes from isospin. While isospin can change at the photon vertex, it must be conserved at the final hadronic vertex. Only with good data on both proton and neutron targets can one hope to disentangle the isoscalar and isovector electromagnetic (EM) couplings of the various N* and Δ* resonances (see Refs. [2,3]), as well as the isospin properties of the non-resonant background amplitudes. The lack of γn→π⁻p and γn→π⁰n data does not allow us to be as confident about the determination of neutron EM couplings relative to those of the proton. For instance, the uncertainties of neutral EM couplings of 4* low-lying N* resonances, $\Delta(nA_{1/2})$ vary between 25 and 140% while charged EM couplings, $\Delta(pA_{1/2})$, vary between 7 and 42%. Some of the N* baryons [N(1675)5/2⁻, for instance] have stronger EM couplings to the neutron relative to the proton, but the parameters are very uncertain [1]. One more unresolved issue relates to the second $P_{11}$, N(1710)1/2⁺. That is not seen in the recent πN PWA [4], contrary to other PWAs used by the PDG14 [1]. A recent brief review of its status is given in Ref. [5]. Obviously, data on the γn→πN reactions are needed to improve the amplitudes and EM couplings.

Additionally, incoherent pion photoproduction on the deuteron is interesting in various aspects of nuclear physics, and particularly, provides information on the elementary reaction on the neutron, i.e., γn→πN. Final-state interaction (FSI) plays a critical role in the state-of-the-art analysis of the γN→πN interaction as extracted from γd→πNN measurements. The FSI was first considered in Refs. [6,7] as responsible for the near-threshold enhancement (Migdal-Watson effect) in the NN mass spectrum of the meson production reaction NN→NNx. In Ref. [8], the FSI amplitude was studied in detail.

## 2 Complete Experiment in Pion Photoproduction

Originally, PWA arose as the technology to determine amplitude of the reaction via fitting scattering data. That is a non-trivial mathematical problem – looking for a solution of ill-posed problem following to Hadamard, Tikhonov *et al.* Resonances appeared as a by-product (bound states objects with definite quantum numbers, mass, lifetime and so on).

There are 4 independent invariant amplitudes for a single pion photoproduction. In order to determine the pion photoproduction amplitude, one has to carry out 8 independent measurements at fixed (W, t) or (E, θ) (the extra observable is necessary to eliminate a sign ambiguity [9]).

---

[a] Corresponding author: igor@gwu.edu

There are 16 non-redundant observables and they are not completely independent from each other, namely 1 unpolarized, $d\sigma/d\Omega$; 3 single polarized, $\Sigma$, $T$, and $P$; 12 double polarized, $E$, $F$, $G$, $H$, $C_x$, $C_z$, $O_x$, $O_z$, $L_x$, $L_z$, $T_x$, and $T_z$ measurements. Additionally, there are 18 triple-polarization asymmetries [9 (9) for linear (circular) polarized beam and 13 of them are non-vanishing] [10,11]. Obviously, the triple-polarization experiments are not really necessary from the theoretical point of view while such measurements will play a critical role to keep systematics under control.

## 3 Neutron Database

Experimental data for neutron-target photoreactions are much less abundant than those utilizing a proton target, constituting only about 15% of the present worldwide known GW SAID database [12]. The existing $\gamma n \rightarrow \pi^- p$ database contains mainly differential cross sections and 15% of which are from polarized measurements. At low to intermediate energies, this lack of neutron-target data is partially compensated by experiments using pion beams, e.g., $\pi^- p \rightarrow \gamma n$, as has been measured, for example, by the Crystal Ball Collaboration at BNL [13] (the GW nuclear physics group was heavily involved in this experiment) for the inverse photon energy E = 285 – 689 MeV and $\theta = 41^0 - 148^0$, where $\theta$ is the inverse production angle of $\pi^-$ in the CM frame. This process is free from complications associated with the deuteron target. However, the disadvantage of using the reaction $\pi^- p \rightarrow \gamma n$ is the 5 to 500 times larger cross sections for $\pi^- p \rightarrow \pi^0 n \rightarrow \gamma \gamma n$, depending on E and $\theta$, which causes a large background, and there were no tagging high flux pion beams.

Figure 1 summarize the available data for single pion photoproduction on the neutron below W = 2.5 GeV. Many high-precision data for the $\gamma n \rightarrow \pi^- p$ and $\gamma n \rightarrow \pi^0 n$ reactions have been measured recently. We applied our GW-ITEP FSI corrections, covering a broad energy range up to E = 2.7 GeV [8], to the CLAS and A2 Collaboration $\gamma d \rightarrow \pi^- pp$ measurements to get elementary cross sections for $\gamma n \rightarrow \pi^- p$ [14,15]. In particular, the new CLAS cross sections have quadrupled the world database for $\gamma n \rightarrow \pi^- p$ above E = 1 GeV. The FSI correction factor for the CLAS (E = 1050 – 2700 MeV and $\theta = 32^0 - 157^0$) and MAMI (E = 410 – 460 MeV and $\theta = 45^0 - 125^0$) kinematics was found to be small, $\Delta\sigma/\sigma < 10\%$.

Obviously, that is not enough to have compatible proton and neutron databases, specifically the energy binning of the CLAS measurements is 50 MeV or, in the worst case, 100 MeV while A2 Collaboration measurements are able to have 2 to 4 MeV binning. The forward direction, which is doable for A2 vs. CLAS, is critical for evaluation of our FSI treatment.

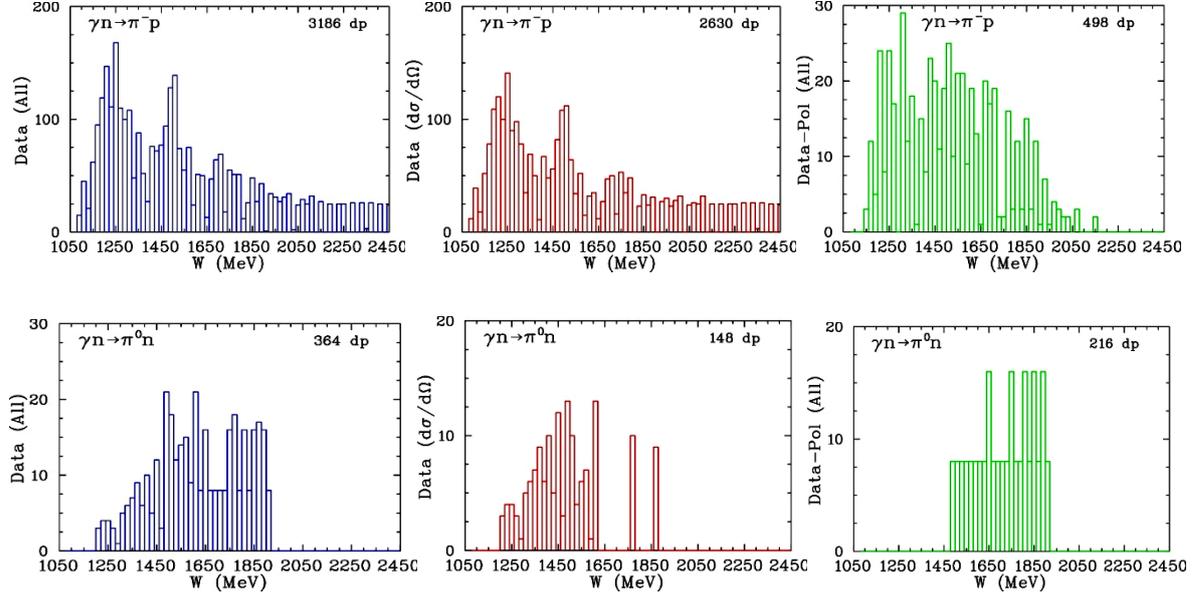

**Figure 1**: Data available for single pion photoproduction off the neutron as a function of CM energy W [12]. The number of data points, dp, is given in the upper right hand side of each subplot. **Row 1:** The first subplot (blue) shows the total amount of $\gamma n \rightarrow \pi^- p$ data available for all observables, the second subplot (red) shows the amount of differential cross-section, $d\sigma/d\Omega$, data available, the third subplot (green) shows the amount of P observables data available. **Row 2:** The first subplot (blue) shows the total amount of $\gamma n \rightarrow \pi^0 n$ data available for all observables, the second subplot (red) shows the amount of $d\sigma/d\Omega$ data available, the third subplot (green) shows the amount of P observables data available.

## 4 Neutron Data from Deuteron Measurements

The determination of the γd→π⁻pp differential cross sections with the FSI, taken into account (including all key diagrams in Fig. **2**), were done, as we did recently [**8**,**14**,**15**], for the CLAS [**14**] and MAMI data [**15**]. The SAID of GW Data Analysis Center (DAC) phenomenological amplitudes for γN→πN [**16**], NN→NN [**17**], and πN→πN [**4**] were used as inputs to calculate the diagrams in Fig. **2**. The Bonn potential (full model) [**18**] was used for the deuteron description. In Ref. [**14**,**15**], we calculated the FSI correction factor $R(E,\theta)$ [Eq. (**1**)] dependent on photon energy, E, and pion production angle in CM frame θ (see details in Refs. [**8**,**14**,**15**]) and fitted recent CLAS and MAMI $d\sigma/d\Omega$ versus the world γN→πN database [**12**] to get new neutron multipoles and determine neutron resonance EM couplings [**14**].

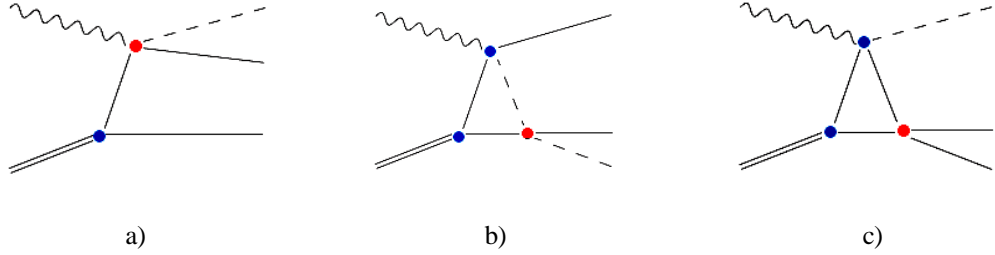

**Figure 2**: Feynman diagrams for the leading components of the γd→π⁻pp amplitude. (**a**) Impulse approximation (IA), (**b**) pp-FSI, and (**c**) πN-FSI. Filled black circles show FSI vertices. Wavy, dashed, solid, and double lines correspond to the photons, pions, nucleons, and deuterons, respectively.

Results of calculations and comparison with the experimental data on the differential cross sections, $d\sigma_{\gamma d}(\theta)/d\Omega$, where Ω and θ are solid and polar angles of outgoing π⁻ in the laboratory frame, respectively, with z-axis along the photon beam for the reaction γd→π⁻pp are given in Fig. **3** (**left**) for a number of the photon energies, E.

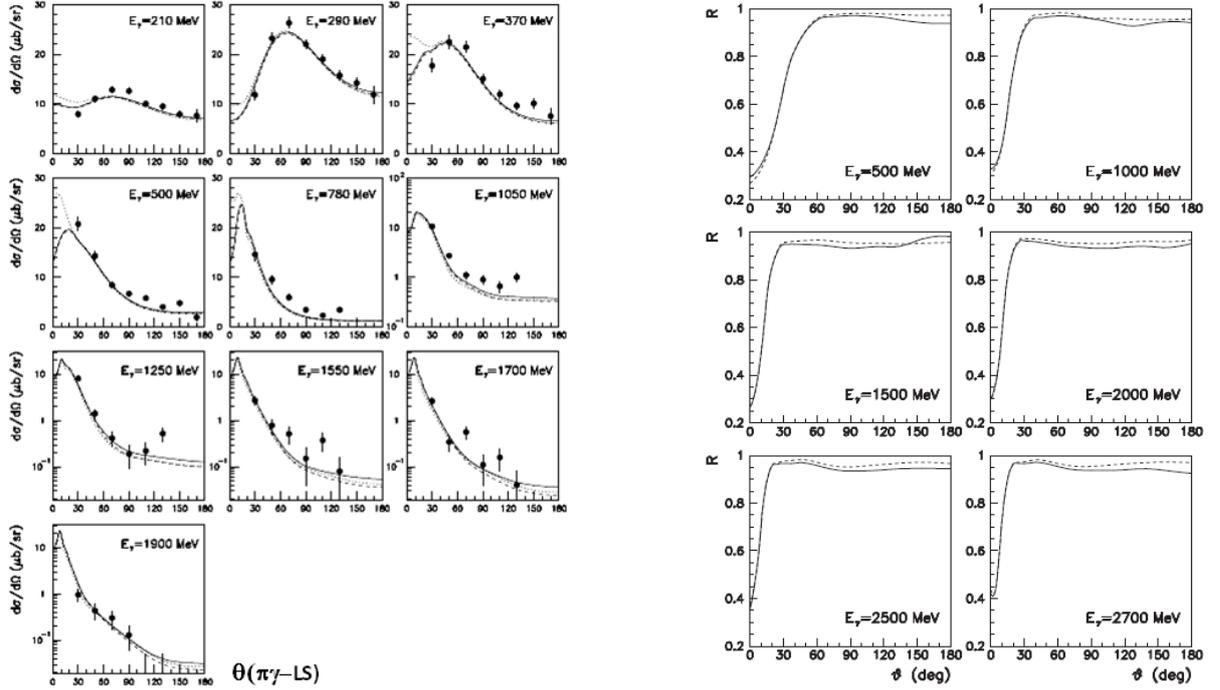

**Figure 3: Left:** The differential cross section, $d\sigma/d\Omega$, of the reaction γd→π⁻pp in the laboratory frame at different values of the photon laboratory energy E < 1900 MeV; θ is the polar angle of the outgoing π⁻. Dotted curves show the contributions from the IA amplitude [Fig. **2**(**a**)]. Successive addition of the NN-FSI [Fig. **2**(**b**)] and πN-FSI [Fig. **2**(**c**)] amplitudes leads to dashed and solid curves, respectively. The filled circles are the data from DESY bubble chamber [**19**]. **Right:** The correction factor R, defined by Eq. (**1**), where θ is the polar angle of the outgoing π⁻ in the rest frame of the pair π⁻ + fast proton. The kinematic cut, $P_p >$ 200 MeV/c, is applied. The solid (dashed) curves are obtained with both πN- and NN-FSI (only NN-FSI) taken into account.

The FSI corrections for the CLAS and MAMI quasi-free kinematics were found to be small, as mentioned above. Our FSI calculations were done [**8**,**14**,**15**] over a broad energy range (threshold to E = 2700 MeV) and for the full angular coverage (θ = 0⁰ – 180⁰). As an illustration, Fig. **3** (**right**) shows the FSI correction factor

$$R(E,\theta) = (d\sigma/d\Omega_{\pi p})/(d\sigma^{IA}/d\Omega_{\pi p}) \qquad (1)$$

for the $\gamma n \rightarrow \pi^- p$ differential cross sections as a function of the pion production angle in the CM ($\pi^- p$) frame, $\theta$, for different energies over the range of the CLAS and MAMI experiments. Overall, the FSI correction factor $R(E,\theta) < 1$, while the effect, i.e., the $(1 - R)$ value, vary from 10% to 30%, depending on the kinematics, and the behavior is very smooth versus pion production angle. We found a sizeable FSI-effect from S-wave part of pp-FSI at small angles. A small but systematic effect $|R - 1| \ll 1$ is found in the large angular region, where it can be estimated in the Glauber approach, except for narrow regions close to $\theta \sim 0^0$ or $\theta \sim 180^0$.

The $\gamma n \rightarrow \pi^0 n$ case is much more complicate vs. $\gamma n \rightarrow \pi^- p$ because $\pi^0 n$ final state can come from both $\gamma n$ and $\gamma p$ initial interactions [20]. The leading diagrams for $\gamma d \rightarrow \pi^0 pn$ are similar as given on Fig, **2**.

## 5 New Neutron Amplitudes and neutron EM Couplings

The solution, SAID GB12 [14], uses the same fitting form as SAID recent SN11 solution [21], which incorporated the neutron-target CLAS $d\sigma/d\Omega$ for $\gamma n \rightarrow \pi^- p$ [14] and GRAAL $\Sigma$s for both $\gamma n \rightarrow \pi^- p$ and $\gamma n \rightarrow \pi^0 n$ [22,23] (Fig. **4**). This fit form was motivated by a multichannel K-matrix approach, with an added phenomenological term proportional to the $\pi N$ reaction cross section.

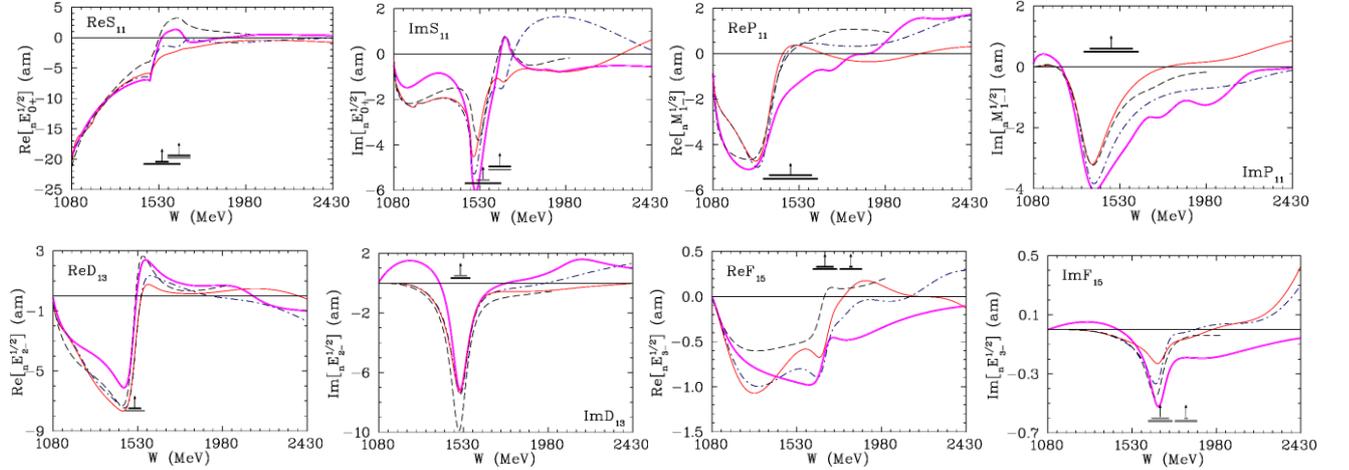

**Figure 4.** Samples of neutron multipoles I = 1/2 and 3/2. Solid (dash-dotted) lines correspond to the SAID GB12 [14] (SN11 [21]) solution. Thick solid (dashed) lines give SAID GZ12 [14] solution (MAID07 [24], which terminates at W = 2 GeV). Vertical arrows indicate mass ($W_R$), and horizontal bars show full, $\Gamma$, and partial, $\Gamma_{\pi N}$, widths of resonances extracted by the Breit-Wigner fit of the $\pi N$ data associated with the SAID solution WI08 [4].

**Table 1.** Neutron helicity amplitudes $A_{1/2}$ and $A_{3/2}$ (in $[(GeV)^{-1/2} \times 10^{-3}]$ units) from the SAID GB12 [14] (first row), previous SAID SN11 [21] (second row), recent BnGa13 by the Bonn-Gatchina group [25] (third row), recent Kent12 by the Kent State Univ. group [26] (forth row), and average values from the PDG14 [1] (fifth row).

| Resonance | $nA_{1/2}$ | Resonance | $nA_{1/2}$ | $nA_{3/2}$ | Ref. |
|---|---|---|---|---|---|
| $N(1535)1/2^-$ | −58± 6 | $N(1520)3/2^-$ | −46± 6 | −115± 5 | SAID GB12 |
| | −60± 3 | | −47± 2 | −125± 2 | SAID SN11 |
| | −93±11 | | −49± 8 | −113±12 | BnGa13 |
| | −49± 3 | | −38± 3 | −101± 4 | Kent12 |
| | −46±27 | | −59± 9 | −139±11 | PDG14 |
| $N(1650)1/2^-$ | −40±10 | $N(1675)5/2^-$ | −58± 2 | −80± 5 | SAID GB12 |
| | −26± 8 | | −42± 2 | −60± 2 | SAID SN11 |
| | 25±20 | | −60± 7 | −88±10 | BnGa13 |
| | 11± 2 | | −40± 4 | −68± 4 | Kent12 |
| | −15±21 | | −43±12 | −58±13 | PDG14 |
| $N(1440)1/2^+$ | 48± 4 | $N(1680)5/2^+$ | 26± 4 | −29± 2 | SAID GB12 |
| | 45±15 | | 50± 4 | −47± 2 | SAID SN11 |
| | 43±12 | | 34± 6 | −44± 9 | BnGa13 |
| | 40± 5 | | 29± 2 | −59± 2 | Kent12 |
| | 40±10 | | 29±10 | −33± 9 | PDG14 |

However, these new CLAS cross sections departed significantly from our predictions at the higher energies, and greatly modified PWA result [14] (Fig. **4**). Recently, the BnGa group reported a neutron EM coupling determination [25]

Dark Matter, Hadron Physics and Fusion Physics

using the CLAS Collaboration $\gamma n \to \pi^- p$ d$\sigma$/d$\Omega$ with our FSI [14] (Table 1). BnGa13 and SAID GB12 used the same (almost) data [14] to fit them while BnGa13 has several new Ad hoc resonances.

Overall: the difference between MAID07 with BnGa13 and SAID GB12 is rather small but resonances may be essentially different (Table 1). The new BnGa13 [25] has some difference vs. GB12 [14], PDG14 [1], for instance, for $N(1535)1/2^-$, $N(1650)1/2^-$, and $N(1680)5/2^+$.

## 6 Work in Progress

At MAMI in March of 2013, we collected deuteron data below E = 800 MeV with 4 MeV energy binning [27] and will have a new experiment below E = 1600 MeV [28] in spring of 2015.

The experimental setup provides close to $4\pi$ sr coverage for outgoing particles [Fig. 5 (**left**)]. The photons from $\pi^0$ decays and charged particles are detected by the CB and TAPS detection system. The energy deposited by charged particles in CB and TAPS is, for the most part, proportional to their kinetic energy, unless they punch through crystals of the spectrometers. Clusters from the final-state neutrons provide information only on their angles. Separation of clusters from neutral particles and charged ones is based on the information from MWPC, PID, and TAPS veto. Separation of positive and negative pions can be based on the identification of the final-state nucleon as either a neutron or a proton. Since cluster energies from charged pions are proportional to their kinetic energy (unless their punch through the crystals), the energy of those clusters can be very low close to reaction threshold.

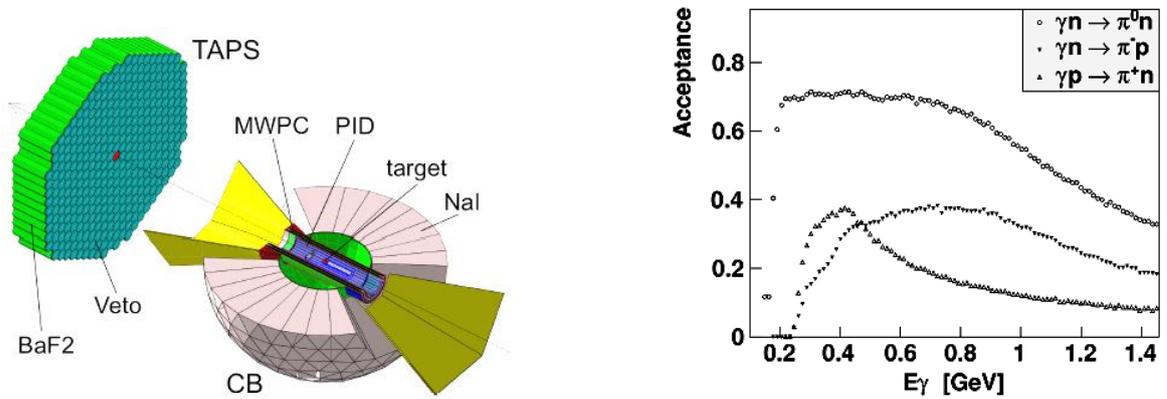

**Figure 5.** **Left:** The A2 detector setup: the Crystal Ball multi-photon spectrometer, with cut-away section showing the inner detectors, and the TAPS forward wall. **Right:** Average acceptance for three reactions $\gamma n \to \pi^0 n$, $\gamma n \to \pi^- p$, and $\gamma p \to \pi^+ n$ with the proposed trigger [28]**.**

Monte Carlo simulations [29], which tracks reaction products through a realistic model of the detector system together with the reconstruction program, is used to calculate acceptance to various channels [Fig. 5 (**left**)]. So to detect the reactions under study with our setup, we have to take data with almost open trigger. In Fig. 5 (**right**), we illustrate the average acceptance obtained for three reactions by requiring the total energy deposited in the CB to be larger than 100 MeV. As seen, acceptance for reaction $\gamma n \to \pi^0 n$ varies from 70% at 0.8 GeV to 30% at 1.5 GeV of the incident-photon energy. Acceptance of reaction $\gamma p \to \pi^+ n$ drops at higher beam energies as charged pions punch through the crystals, and the energy of the neutron cluster does not reflect its kinetic energy. Reaction $\gamma n \to \pi^- p$ above 0.8 GeV has an acceptance that is better than that for $\gamma p \to \pi^+ n$ as the energy and angles of the cluster from the outgoing proton can be used to reconstruct the reaction kinematics.

We are going to use our FSI technology to apply for the upcoming JLab CLAS (g13 run period) d$\sigma$/d$\Omega$ for $\gamma n \to \pi^- p$ covering E = 400 − 2500 MeV and $\theta = 18^0 - 152^0$ [30]. This data set will bring about 12k new measurements which quadruple the world $\gamma n \to \pi^- p$ database. The ELPH facility at Tohoku Univ. will bring new d$\sigma$/d$\Omega$ for $\gamma n \to \pi^0 n$ below E = 1200 MeV [31].

## 7 Polarized Measurements

The difference between previous pion photoproduction and new polarized measurements may result in significant changes in the neutron EM couplings. The strategy for the FSI calculations addressed to $\Sigma(\gamma d \to \pi^- pp)$ to get $\Sigma(\gamma n \to \pi^- p)$ is under consideration now. Our preliminary conclusion is that the FSI correction for the $\Sigma$-beam asymmetry is small far away from the pion threshold.

The SAID SP09 solution [32] is consistent with the GRAAL $\Sigma$-data for $\gamma n \to \pi^- p$ in the forward angular region where previous results constrained the fit (Fig. 6). In the backward region and at energies above 1100 MeV, the agreement becomes satisfactory only after inclusion of GRAAL data. The MAID2007 [24] solution agrees with GRAAL data in the forward region. Both SAID-SP09 and MAID2007 results exhibit structures not seen in the data and which explain the poor $\chi^2$ for both cases.

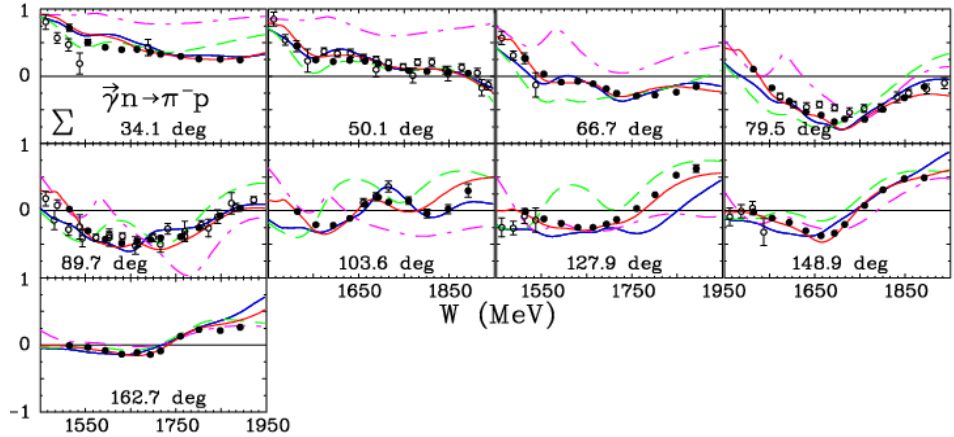

**Figure 6.** The Σ-beam polarization asymmetries for γn→π⁻p versus CM energy W. The pion CM production angle is shown. Solid red (solid blue) lines correspond to the SAID MA09 [22] (SP09 [32]) solution. Green dashed (magenta dot-dashed) lines give the MAID2007 [24] (DMT [33]) predictions. Experimental data are from the GRAAL [22] (filled circles) and previous measurements [12] (open circles). The plotted points from previously published experimental data are those data points within 4 MeV of the photon energy indicated on each panel. Plotted uncertainties are statistical. The MA09 includes in its database the GRAAL asymmetries for γn→π⁻p [22] and γn→π⁰n [23] reactions. SP09 and MAID2007 do not include these Σ data.

The upcoming JLab CLAS (g13 run period) Σ-data for γn→π⁻p, covering E = 910 − 2394 MeV and θ = 18⁰ − 131⁰, [33] have small disagreement with recent GRAAL measurements [22].

Previous γn→π⁻p measurements provided a better constraint vs. γn→π⁰n case. 216 new GRAAL Σs for γn→π⁰n are 60% of the world π⁰n data (Fig. 7). For the GRAAL γn→π⁰n (γn→π⁻p), the SAID SP09 had $\chi^2$/data = 223 (89) while SAID MA09 got $\chi^2$/data = 3.1 (4.9). SAID MA09 included both GRAAL Σ-measurements [22,23].

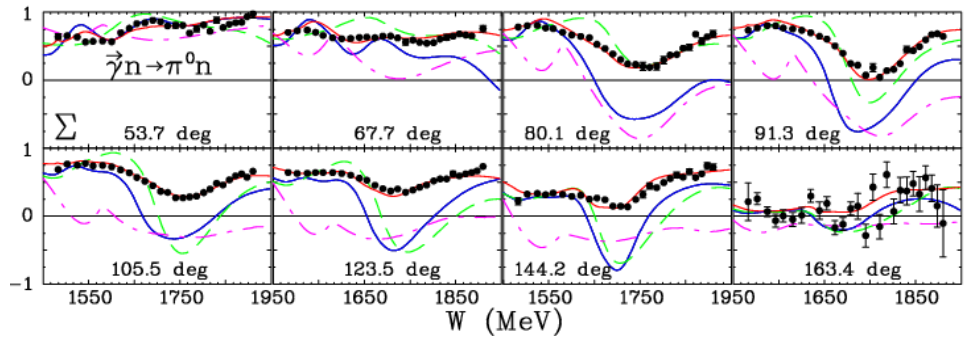

**Figure 7.** The Σ-beam polarization asymmetries for γn→π⁰n versus CM energy W. Notation as in Fig. 6.

## 8 Summary for Neutron Study

- The differential cross section for the processes γn→π⁻p was extracted from new CLAS and MAMI-B measurements accounting for Fermi motion effects in the IA as well as NN- and πN-FSI effects beyond the IA.
- Consequential calculations of the FSI corrections, as developed by the GW-ITEP Collaboration, was applied.
- New cross sections departed significantly from our predictions, at the higher energies, and greatly modified the fit result.
- New γn→π⁻p and γn→π⁰n data will provide a critical constraint on the determination of the multipoles and EM couplings of low-lying baryon resonances using the PWA and coupled channel techniques.
- Polarized measurements at JLab/JLab12, MAMI, SPring-8, ELSA, and ELPH will help to bring more physics in.
- FSI corrections need to apply.

**Acknowledgements:** This material is based upon work supported by the U.S. Department of Energy, Office of Science, Office of Nuclear Physics, under Award Number DE-FG02-99-ER41110. AK, MM, VK, and VT thank the Institute for Kernphysik at Mainz where part of this work was performed for hospitality and support.